\documentclass[journal]{IEEEtran} 
\usepackage[utf8]{inputenc}

\usepackage{tabularx}
\usepackage{booktabs}
\usepackage{multicol}
\usepackage{multirow}
\usepackage{float}

\usepackage{amsmath,amssymb}

\usepackage{graphicx}
\usepackage{xcolor}
\usepackage{qcircuit}

\usepackage{hyperref}
\usepackage{cite}

\title{Accelerating Parameter Initialization in Quantum Chemical Simulations via LSTM-FC-VQE}

\author{
Ran-Yu Chang\IEEEauthorrefmark{1}, 
Yu-Cheng Lin\IEEEauthorrefmark{2}, 
Pei-Che Hsu\IEEEauthorrefmark{3}, 
Tsung-Wei Huang\IEEEauthorrefmark{4}, 
En-Jui Kuo\IEEEauthorrefmark{5}%
\thanks{Ran-Yu Chang \IEEEauthorrefmark{1} and  Yu-Cheng Lin\IEEEauthorrefmark{2} contributed equally to this work.} \\
\IEEEauthorrefmark{1}Arete Honors Program, National Yang Ming Chiao Tung University, Hsinchu, Taiwan \\
\IEEEauthorrefmark{2}Quantum Technology Office, National Yang Ming Chiao Tung University, Hsinchu, Taiwan \\
\IEEEauthorrefmark{3}Department of Electrical and Computer Engineering, National Yang Ming Chiao Tung University, Hsinchu, Taiwan \\
\IEEEauthorrefmark{4}Department of Intelligent Computing and Big Data, Chung Yuan Christian University, Taoyuan, Taiwan \\
\IEEEauthorrefmark{5}Department of Electrophysics, National Yang Ming Chiao Tung University, Hsinchu, Taiwan \\
\texttt{leo07010@gmail.com, nagi30912@gmail.com, 25425108a@gmail.com, tsungwei@cycu.edu.tw, kuoenjui@nycu.edu.tw}
}

\begin{document}

\maketitle
\IEEEpeerreviewmaketitle

\maketitle
\begin{abstract}
We present a meta-learning framework that leverages Long Short-Term Memory (LSTM) neural networks to accelerate parameter initialization in quantum chemical simulations using the Variational Quantum Eigensolver (VQE). By training the LSTM on optimized parameters from small molecules, the model learns to predict high-quality initializations for larger systems, reducing the number of required VQE iterations. Our enhanced LSTM-FC-VQE architecture introduces a fully connected layer, improving adaptability across molecules with varying parameter sizes. Experimental results show that our approach achieves faster convergence and lower energy errors than traditional initialization, demonstrating its practical potential for efficient quantum simulations in the NISQ era.
\end{abstract}


\section{Introduction}
Quantum computing has shown great potential in the
fields of chemistry and materials science, particularly
through the application of the Variational Quantum
Eigensolver (VQE) in solving molecular ground-state
energies \cite{Peruzzo2014}. In the current Noisy Intermediate-Scale
Quantum (NISQ) era \cite{NISQ}, hardware constraints and the
presence of noise often require a large number of iterations
or measurements for acceptable convergence, especially
when handling larger molecular systems, where classical
simulation becomes exponentially expensive \cite{Cao2019}.

VQE is a famous hybrid quantum-classical  algorithm \cite{Peruzzo2014,OpenFermion} that employs a parameterized quantum circuits  (PQC) to generate quantum states, with a classical optimizer iteratively adjusting parameters to minimize the expectation value of target Hamiltonian. While VQE mitigates the exponential scaling of computational resources \cite{Cao2019}, its performance is constrained by hardware noise, short coherence times, and limited circuit depth \cite{NISQ,Kandala2017}. Additionally, suboptimal parameter initialization and naive local optimizers can result in slow convergence, as they may become trapped in local minima or require extensive function evaluations, especially when initialization relies on random or heuristic methods \cite{Kandala2017}.

VQE has gained traction not only in molecular energy
calculations but also in broader applications such as quantum
magnetism, materials property predictions, and even certain
combinatorial optimization tasks \cite{s2023efficientvqeapproachaccurate,Choudhary_2021}. Furthermore, ongoing research
has explored employing VQE for simulating chemical reaction
pathways\cite{Chee2022}, designing novel catalysts, and investigating strongly
correlated electron systems --- underscoring its promise in
advancing both fundamental science and industrial applications
\cite{Ref14}. 

To mitigate these issues, existing approaches have explored
(a) random multiple-start seeds and then selecting the best local optimum \cite{Ref15},
(b) heuristic methods such as adiabatic state preparation \cite{10.1145/3479197}, or
(c) hardware-efficient ans\"atze combined with gradient-based optimizers \cite{OpenFermion}.
While these strategies can partially address the problem, they
do not guarantee a globally efficient initialization for every new system.
Therefore, we propose utilizing machine learning techniques to train a model capable of learning how to predict potential initialization parameter positions when presented with a new wavefunction, ensuring rapid convergence to experimental requirements, a meta-learning  framework \cite{finn2017model,vilalta2002perspective} such as an LSTM  \cite{LongShortTermMemory1997} optimizer has emerged as a promising alternative, offering a more systematicway to reuse knowledge across different Hamiltonians and molecule
sizes.

To address the challenges associated with parameter initialization and optimization in VQE, researchers have proposed the Meta-VQE framework, which integrates meta-learning techniques to enhance initialization strategies and improve convergence. The Meta-VQE method divides quantum circuits into encoding and processing layers, enabling the algorithm to learn the energy profiles of parameterized Hamiltonians with a limited set of training data. This approach reduces reliance on random initialization, effectively mitigates issues such as barren plateaus, and enhances the adaptability of VQE across different Hamiltonians \cite{MetaVQE2021,OptimizedMetaVQE2022}.

Building on this foundation, optimized Meta-VQE strategies introduce refined cost functions and flexible circuit architectures to address noise and fixed circuit structure limitations. These enhancements improve the algorithm's ability to learn energy trends and provide more accurate estimations for equilibrium bond lengths \cite{OptimizedMetaVQE2022}. Moreover, non-linear encodings, such as Gaussian-based encodings, have been employed to better approximate complex energy landscapes and improve convergence rates \cite{MetaVQE2021}.

Further advancements by Verdon et al. introduced the "learning to learn" paradigm using LSTM networks, leveraging their sequential memory capabilities to train predictive models for wavefunction parameters. This framework demonstrated significant potential in providing high-quality initialization parameters for VQE, accelerating convergence while maintaining accuracy \cite{Verdon2019,MetaVQE2021}. These contributions collectively position Meta-VQE as a scalable and adaptable method for advancing quantum simulation and quantum chemistry applications under NISQ constraints.

In the context of quantum chemistry, the search space
is often vast due to the exponentially large electronic
configuration. Efficient initialization becomes crucial \cite{McArdle2020}.
By training an LSTM on smaller molecular systems, it is
possible to learn a mapping from molecular features (or
Hamiltonian coefficients) to near-optimal circuit parameters, thus reducing the total iteration count when tackling
bigger molecules. Furthermore, this approach helps
lower the overhead of quantum-classical communication,
which is especially beneficial under NISQ constraints \cite{Harrigan2021}.

LSTM-based meta-learning has
proven effective in navigating high-dimensional parameter
spaces more efficiently \cite{Finn2017, Verdon2019, tsai2020learning, tsai2022path}. Hence, combining VQE
with an LSTM meta-learner allows for capturing complex
relationships between molecular descriptors and optimized
parameter sets, with some studies observing an order-of-
magnitude reduction in iterations for related QAOA/VQE
tasks \cite{Verdon2019}. Building on these insights, our work further
explores the scalability and adaptability of this technique
for more intricate Hamiltonians, aiming to accelerate
VQE-based simulations of large molecules within the
constraints of current quantum hardware.

In this work, we highlight a key challenge in applying VQE to simulate molecular ground-state energies: the number of Ansatz parameters required by VQE varies depending on the molecular system. This leads to an inconsistency in parameter dimensions, making it fundamentally different from typical "learning to learn" approaches that assume a fixed number of parameters for simpler systems. To address this, we propose a novel method that enables the LSTM to handle variable-dimensional input parameters, ensuring flexibility and adaptability in the VQE-LSTM framework. This innovation allows the VQE-LSTM approach to accommodate a broader range of molecular systems, significantly enhancing its versatility and practical applicability in quantum chemical simulations.

\section{Method and Results}
All quantum algorithms and machine learning models in this work were implemented using the following Python packages and versions: TensorFlow 2.19.0\cite{tensorflow2015}, Keras 3.9.0\cite{chollet2015keras}, PennyLane 0.40.0\cite{antipov2022pennylane}, SciPy 1.15.2\cite{roy2023scipy}, NumPy 2.0.2\cite{klein2019numpy}, and Matplotlib 3.10.1\cite{white2020matplotlib}. The experiments were executed with Python 3.10.16.
Further implementation and training details are provided in the Appendix.
\subsection{Variational Quantum Eigensolver}
In quantum chemistry, the VQE is a hybrid quantum-classical algorithm designed to estimate the ground-state energy of molecular systems. At its core, VQE leverages a parameterized quantum state , known as the \textit{ansatz}, to approximate the ground state of a Hamiltonian . Common choices of ansatz include the Hardware-Efficient Ansatz (HEA)\cite{kandala2017hardware} and the Unitary Coupled Cluster (UCC) ansatz \cite{kutzelnigg1991structure,taube2006new} . The algorithm aims to minimize the expectation value

\begin{equation}
\begin{aligned}
E(\boldsymbol{\theta}) = \langle \psi(\boldsymbol{\theta}) | H | \psi(\boldsymbol{\theta}) \rangle,
\end{aligned}
\end{equation}

using classical optimizers. In quantum chemistry, the electronic Hamiltonian under second quantization can be expressed as:

\begin{equation}
\begin{aligned}
H = \sum_{p,q} h_{pq} a_p^\dagger a_q + \frac{1}{2} \sum_{p,q,r,s} h_{pqrs} a_p^\dagger a_q^\dagger a_r a_s,
\end{aligned}
\end{equation}

where \(h_{pq}\) and \(h_{pqrs}\) represent one- and two-electron integrals, and \(a^\dagger, a\) are fermionic creation and annihilation operators. These integrals are typically obtained via self-consistent field (SCF) \cite{hartree1928wave} calculations prior to VQE optimization.

As shown in Figure 1, once the molecular Hamiltonian is obtained, the procedure begins with the design of an ansatz. Next, a set of initial parameters (angles) for the ansatz is randomly selected. The expectation value of the Hamiltonian is then evaluated by measuring its decomposition into Pauli operators. Finally, a classical optimizer is employed to iteratively update the ansatz parameters until the estimated energy converges.
\begin{figure}[h]
    \centering
\includegraphics[width=\linewidth]{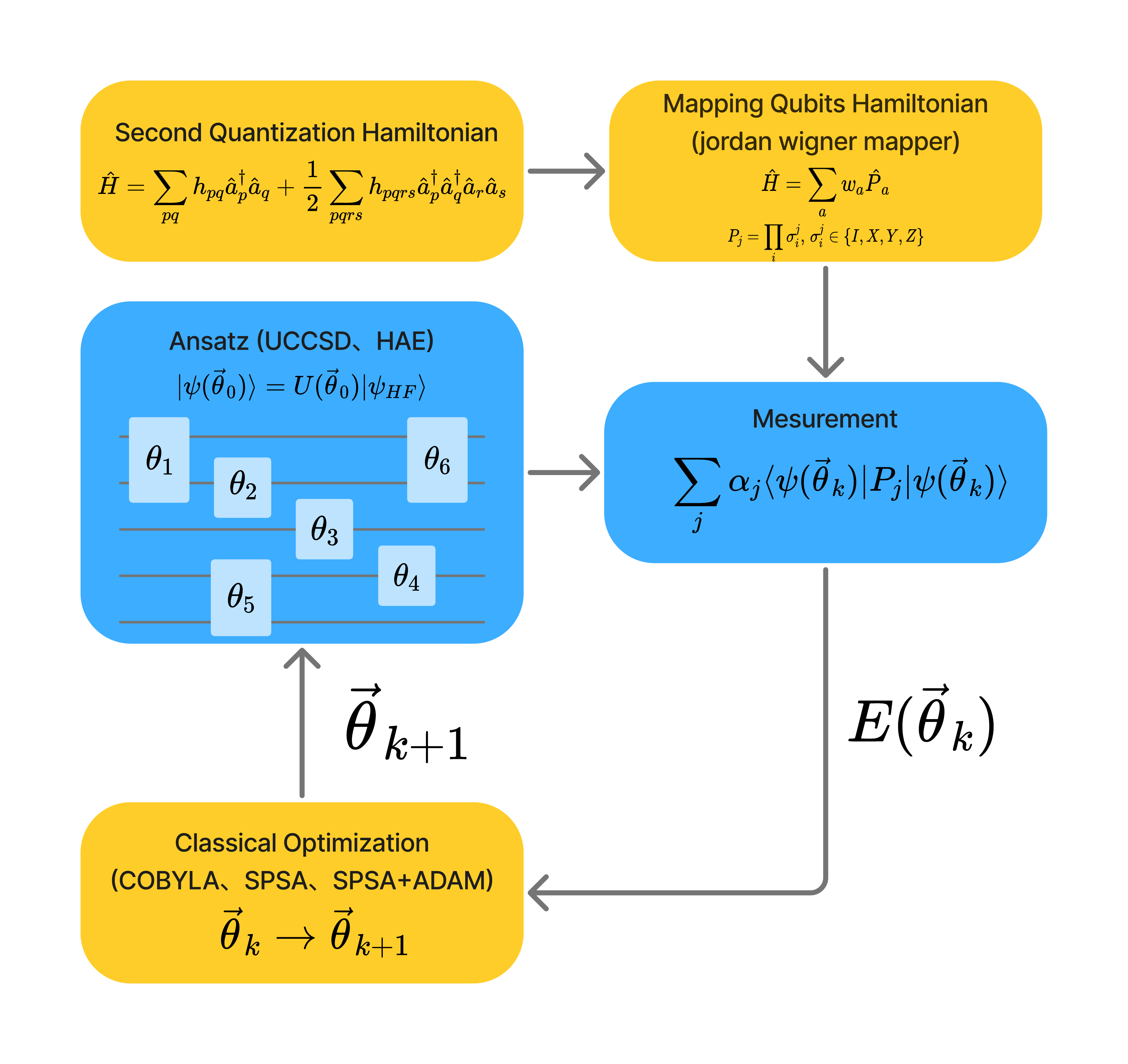}
    \caption{Schematic of the VQE algorithm. The algorithm alternates between preparing quantum states using a parameterized circuit and estimating the energy via quantum measurements. A classical optimizer then updates the circuit parameters to minimize the energy. This iterative process continues until convergence, yielding the ground-state energy and wavefunction of the target Hamiltonian.}
    \label{fig:enter-label}
\end{figure}

\subsection{Long Short-Term Memory}

LSTM networks are a specialized type of recurrent neural network (RNN) \cite{elman1990finding} designed to capture long-range dependencies and overcome the vanishing gradient problem inherent in standard RNNs. Originally proposed by Hochreiter and Schmidhuber in 1997\cite{LongShortTermMemory1997}, LSTM architectures have since become foundational in a wide range of sequence modeling tasks, including natural language processing, time-series forecasting, and sequential decision-making.

An LSTM unit maintains a memory cell $\mathbf{c}_t$ that is updated through carefully designed gates: the input gate $\mathbf{i}_t$, forget gate $\mathbf{f}_t$, and output gate $\mathbf{o}_t$. These gates regulate the flow of information into and out of the memory cell, defined mathematically as follows:

\begin{equation}
\begin{aligned}
\mathbf{f}_t &= \sigma(\mathbf{W}_f [\mathbf{h}_{t-1}, \mathbf{x}_t] + \mathbf{b}_f), \\
\mathbf{i}_t &= \sigma(\mathbf{W}_i [\mathbf{h}_{t-1}, \mathbf{x}_t] + \mathbf{b}_i), \\
\tilde{\mathbf{c}}_t &= \tanh(\mathbf{W}_c [\mathbf{h}_{t-1}, \mathbf{x}_t] + \mathbf{b}_c), \\
\mathbf{c}_t &= \mathbf{f}_t \odot \mathbf{c}_{t-1} + \mathbf{i}_t \odot \tilde{\mathbf{c}}_t, \\
\mathbf{o}_t &= \sigma(\mathbf{W}_o [\mathbf{h}_{t-1}, \mathbf{x}_t] + \mathbf{b}_o), \\
\mathbf{h}_t &= \mathbf{o}_t \odot \tanh(\mathbf{c}_t)
\end{aligned}
\end{equation}

where $\mathbf{x}_t$ is the input vector at time $t$, $\mathbf{h}_t$ is the hidden state, and $\sigma$ denotes the sigmoid activation function. The element-wise product $\odot$ controls how much past memory is retained and how much new information is incorporated.

In the context of quantum machine learning, LSTMs are particularly well-suited for tasks where the input data represents structured quantum information, such as Hamiltonians, excitation integrals, or parameterized quantum circuits. Their ability to process variable-length input and capture complex dependencies makes them a strong candidate for learning initialization heuristics for VQE, as explored in our work.

\subsection{LSTM-VQE}
LSTM-VQE method based on the RNN-QNN framework proposed by Verdon et al\cite{Verdon2019}., effectively addressing the varying dimensionality of parameters across different chemical molecular systems. The method integrates LSTM neural networks with Parameterized Quantum Circuits (PQC), leveraging the predictive capabilities of neural networks to achieve quantum-enhanced learning.

Since molecular Hamiltonians exhibit variability in the dimensions of their one-electron and two-electron integrals, traditional LSTM networks cannot directly handle such discrepancies. To address this issue, we utilize a padding strategy to standardize integral data from different molecules into a uniform dimension, ensuring consistent input sizes for the LSTM. The maximum dimension used as input to the LSTM is defined mathematically as:

\begin{equation}
\begin{aligned}
\mathrm{LSTM_{params}} &= \{p_1, \cdots, p_N\}
\end{aligned}
\end{equation}

\begin{equation}
\begin{aligned}
p_{s}&=p_{1:N/2}\\
p_{d}&=p_{N/2+1:N}
\end{aligned}
\end{equation}

We define the padding operation as:
\begin{equation}
\theta_{s/d} =
\begin{cases}
[\theta^{max,s/d}_{1}, x_2, \dots, x_N, 0, \dots, 0], & \text{if } N < M, \\
[x_1, x_2, \dots, x_M], & \text{if } N \ge M.
\end{cases}
\end{equation}

Here, if the original parameter dimension $N$ is smaller than $M$, we pad the remaining $M - N$ positions with zeros (or other suitable values). If $N$ exceeds $M$, we truncate and only keep the first $M$ components.

Let the LSTM network process $\tilde{\mathbf{x}}$ to produce an output vector $\hat{\mathbf{y}} \in \mathbb{R}^M$:
\begin{equation}
\hat{\mathbf{y}} = \mathrm{LSTM}\bigl(\tilde{\mathbf{x}}\bigr).
\end{equation}

Finally, we retain only the first $N$ dimensions of the output, denoted by $\mathbf{y} \in \mathbb{R}^N$:
\begin{equation}
\mathbf{y} = [\hat{y}_1, \hat{y}_2, \dots, \hat{y}_N].
\end{equation}

Through this \emph{pad-then-truncate} procedure, the LSTM can maintain a fixed weight dimension of $M$ internally, while still accommodating variable input lengths $N$ from different training samples.

Operationally, single-electron and double-electron integrals for each molecule are initially computed using SCF methods and subsequently expanded into unified parameter vectors. For molecules with fewer parameters, padding is applied to match the predefined dimension. These standardized vectors are then fed into the LSTM model, which predicts suitable initial parameters for the VQE ansatz (e.g., HEA or UCC). A quantum computer calculates the energy expectation value using these predicted parameters, and a classical optimizer (such as ADAM \cite{kingma2014adam} or SGD \cite{goodfellow2016deep}) iteratively refines the parameters until convergence is achieved.

The loss function is defined as a weighted average, with each layer of the RNN providing an energy result. The final loss function is expressed as:
\begin{equation}
L = \frac{1}{T}\sum_t^T (0.1 \times t) E_t(\boldsymbol{\theta}),
\end{equation}
where
\begin{equation}
E_t(\boldsymbol{\theta}) = \langle \psi(\boldsymbol{\theta}_{t}) | H | \psi(\boldsymbol{\theta}_{t}) \rangle.
\end{equation}

\subsection{LSTM-FC-VQE (Fully Connected Extension)}

While the LSTM-VQE framework can effectively standardize molecular features through padding strategies, the original architecture still faces a major limitation. Since the number of parameters required by a VQE ansatz depends on the number of qubits and the chosen circuit structure, the LSTM output vector $\phi_t$ must be projected or truncated to match the parameter shape required for each molecule. This leads to dispersed learning representations, increased computational cost, longer training times, and a potential loss of critical parameter patterns—ultimately degrading both prediction accuracy and model generalization.

To address this issue, we propose the \textbf{LSTM-FC-VQE} architecture, which appends a molecule-specific fully connected (FC) layer after the LSTM output. Mathematically, this can be formulated as:

\begin{equation}
\theta_t = \mathrm{FC}^{(m)}\bigl( \phi_t \bigr),
\end{equation}

where $\mathrm{FC}^{(m)}$ denotes the FC layer corresponding to molecule $m$ (or more generally, the combination of ansatz structure and qubit layout). This FC layer maps the fixed-dimensional LSTM output $\phi_t \in \mathbb{R}^M$ to the required parameter dimension for the specific ansatz:

\begin{equation}
\theta_t \in \mathbb{R}^{N_m}, \quad N_m = \mathrm{dim}\left( \theta^{\text{Ansatz}_m} \right).
\end{equation}

The FC layer serves two primary purposes:
\begin{itemize}
\item It decouples the LSTM's output dimension from the ansatz-specific parameter count, allowing the core LSTM architecture to remain fixed across molecules.
\item It acts as a projection head that conditions the latent representation $\phi_t$ on the target molecule and its ansatz structure.
\end{itemize}

In practice, each ansatz type is associated with a trainable FC layer. During training, the model jointly learns transferable latent representations through the LSTM and adapts them via the FC layer to match the parameter requirements of each molecule.

\begin{figure}[h]
\centering
\includegraphics[width=\linewidth]{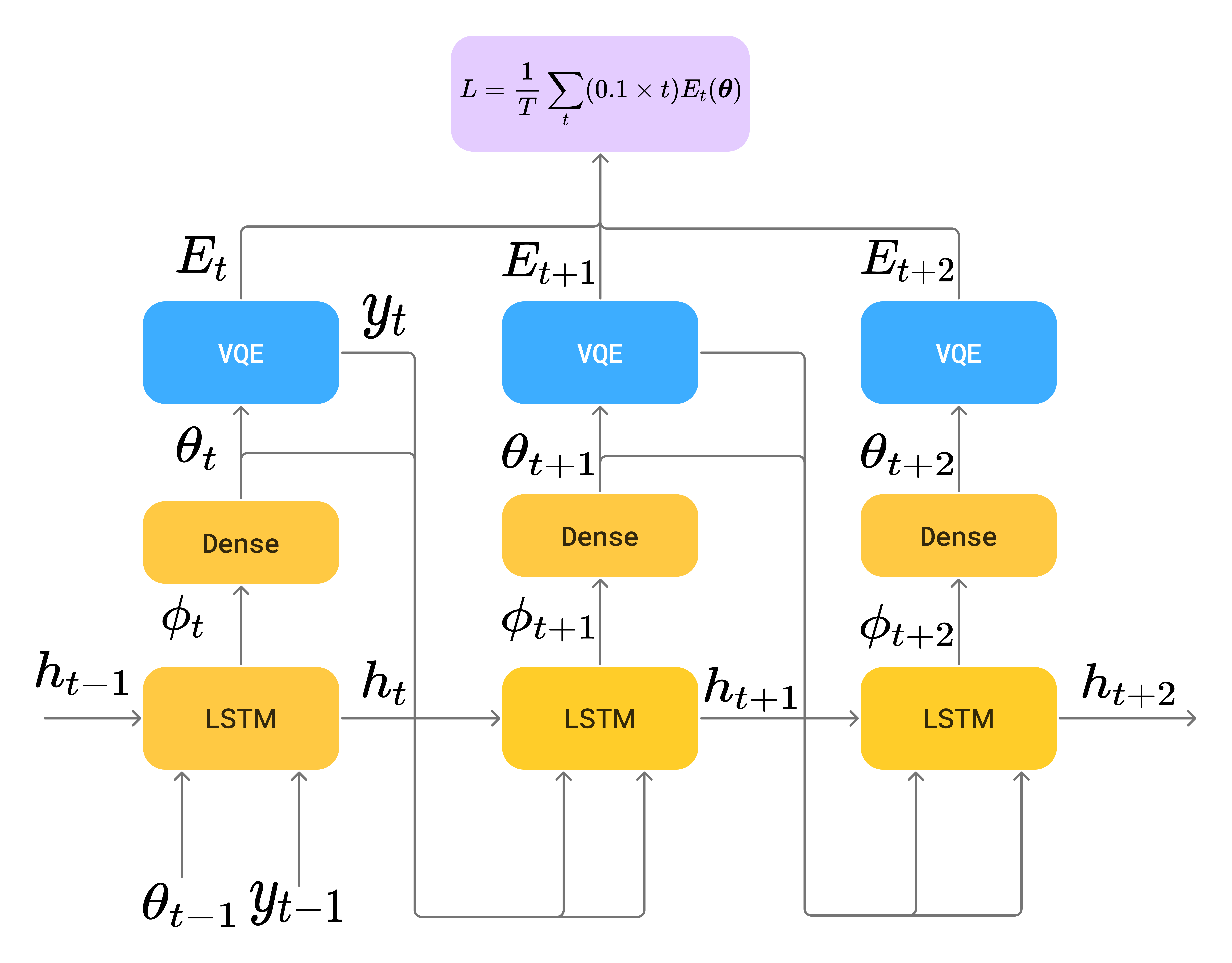}
\caption{LSTM-FC-VQE Framework: At each time step $t$, the LSTM receives $\theta_{t-1}$ and $y_{t-1}$ as inputs and outputs a latent representation $\phi_t$, which is transformed into VQE parameters $\theta_t$ via a fully connected layer. VQE evaluates energy $E_t$, and the final loss is a weighted sum across all $T$ steps.}
\label{fig:LSTM-FC-VQE}
\end{figure}

This architecture improves both generalization and expressiveness. Unlike conventional meta-VQE methods that rely on interpolation or random initialization, LSTM-FC-VQE integrates a shared latent space with molecule-conditioned adaptation, enabling it to efficiently generate high-quality initial parameters and accelerate VQE convergence while preserving chemical accuracy.

\section{Result}
Table~\ref{tab:h4_initialization} presents the number of optimization iterations and the corresponding final energy errors (in units of mHa) for the H\textsubscript{4} molecule under four different parameter initialization strategies \cite{uccsd-vqe,intial}: Random, All Zero, LSTM, and LSTM with a fully connected layer. In the Random scheme, each variational parameter is initialized independently by drawing from a uniform distribution over the interval $[0,1]$. All Zero refers to initializing all parameters to zero, which in UCCSD/VQE corresponds to starting from the Hartree–Fock reference state. Each initialization strategy was evaluated under both constant and decaying learning rate schedules, using SGD and Adam optimizers.

Romero et al. \cite{intial} employed the Unitary Coupled Cluster (UCC) ansatz within the VQE framework to compute molecular energies, achieving chemical accuracy of 4--6 kcal/mol (approximately 6.37--9.56 mHa, using the conversion 1 kcal/mol $\approx$ 1.593 mHa) for the linear H\textsubscript{4} molecule. However, their method required over 400 optimization iterations to converge. In contrast, our fully connected LSTM model, trained with a constant learning rate and the Adam optimizer, achieves a significantly lower error of 0.048 mHa within only 86 iterations—well below the threshold of chemical accuracy—demonstrating both superior efficiency and precision, without the need for circuit optimization.

Notably, traditional VQE approaches with random or zero parameter initialization require substantially more iterations to converge. For example, random initialization followed by stochastic gradient descent (SGD) demands 290 iterations to reach an error of 28.53 mHa. Even with LSTM initialization alone, 357 iterations are needed, suggesting that the base LSTM architecture still struggles to generalize effectively without architectural enhancements.

These findings highlight the advantage of incorporating a fully connected layer into the LSTM-VQE architecture. This modification enables more accurate parameter prediction and improved convergence behavior, particularly when coupled with the Adam optimizer and an appropriately tuned learning rate schedule.

\begin{table}[H]
\centering
\caption{Comparison of optimization iterations and final energy error (mHa) for H$_4$ with different parameter initialization strategies.}
\label{tab:h4_initialization}
\resizebox{\columnwidth}{!}{%
\begin{tabular}{c|c|cc|cc}
\toprule
\multirow{2}{*}{Initialization Method} & \multirow{2}{*}{Learning Rate} & \multicolumn{2}{c|}{SGD} & \multicolumn{2}{c}{Adam} \\
\cmidrule(lr){3-4} \cmidrule(lr){5-6}
& & Iterations & Error (mHa) & Iterations & Error (mHa) \\
\midrule
\multirow{2}{*}{Random} 
         & Constant & 290 & 28.53 & 96 & 28.54 \\
         & Decay & 102 & 28.46 & $\geq$ 501 & 15.92 \\
\midrule
\multirow{2}{*}{All Zero} 
         & Constant & 52 & 28.46 & 65 & 28.48 \\
         & Decay & 97 & 28.46 & 101 & 18.11 \\
\midrule
\multirow{2}{*}{LSTM} 
         & Constant & 357 & 28.53 & 94 & 28.50 \\
         & Decay & 102 & 28.46 & $\geq$ 501 & 13.86 \\
\midrule
\multirow{2}{*}{LSTM with Fully Connected} 
         & Constant & 333 & 0.16 & 86 & 0.05 \\
         & Decay & 222 & 0.14 & 256 & 17.41 \\
\bottomrule
\end{tabular}%
}
\end{table}

Given that the LSTM-FC-VQE model demonstrated the best performance among all tested approaches, we further assessed its accuracy by analyzing the potential energy curve of the H\textsubscript{4} molecule across a range of interatomic distances.

\begin{figure}[htbp]
    \centering
    \includegraphics[width=1\linewidth]{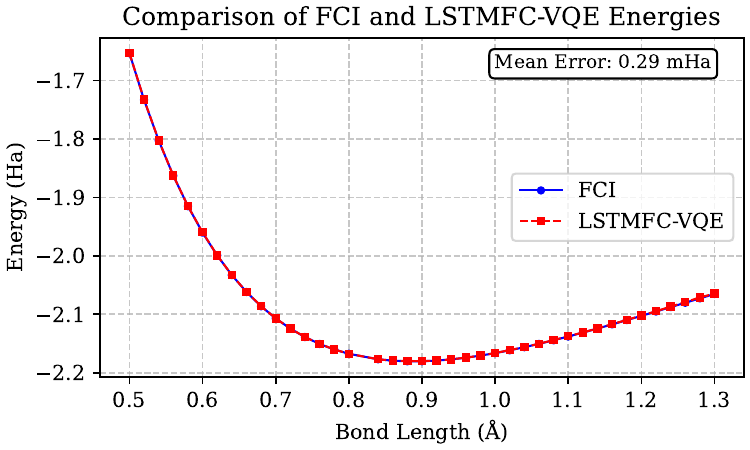}
    \caption{Potential energy curve of the H\textsubscript{4} molecule as a function of bond length.}
    \label{fig:h4 energy}
\end{figure}

As illustrated in Figure~\ref{fig:h4 energy}, the energy curve predicted by the LSTM-FC-VQE model (green dashed line) closely tracks the reference FCI results (blue solid line), exhibiting excellent agreement throughout the entire bond length range. Notably, the model accurately reproduces key features of the potential energy surface, including the location of the energy minimum and the curvature surrounding it.

These results demonstrate that the LSTM-FC-VQE model is capable not only of achieving low error in single-point energy predictions, but also of capturing the broader physical trends in molecular energy landscapes. Such robustness is essential for practical quantum simulations involving bond stretching, molecular dissociation, or reaction pathway analysis.

\begin{figure}[htbp] \centering \includegraphics[width=1\linewidth]{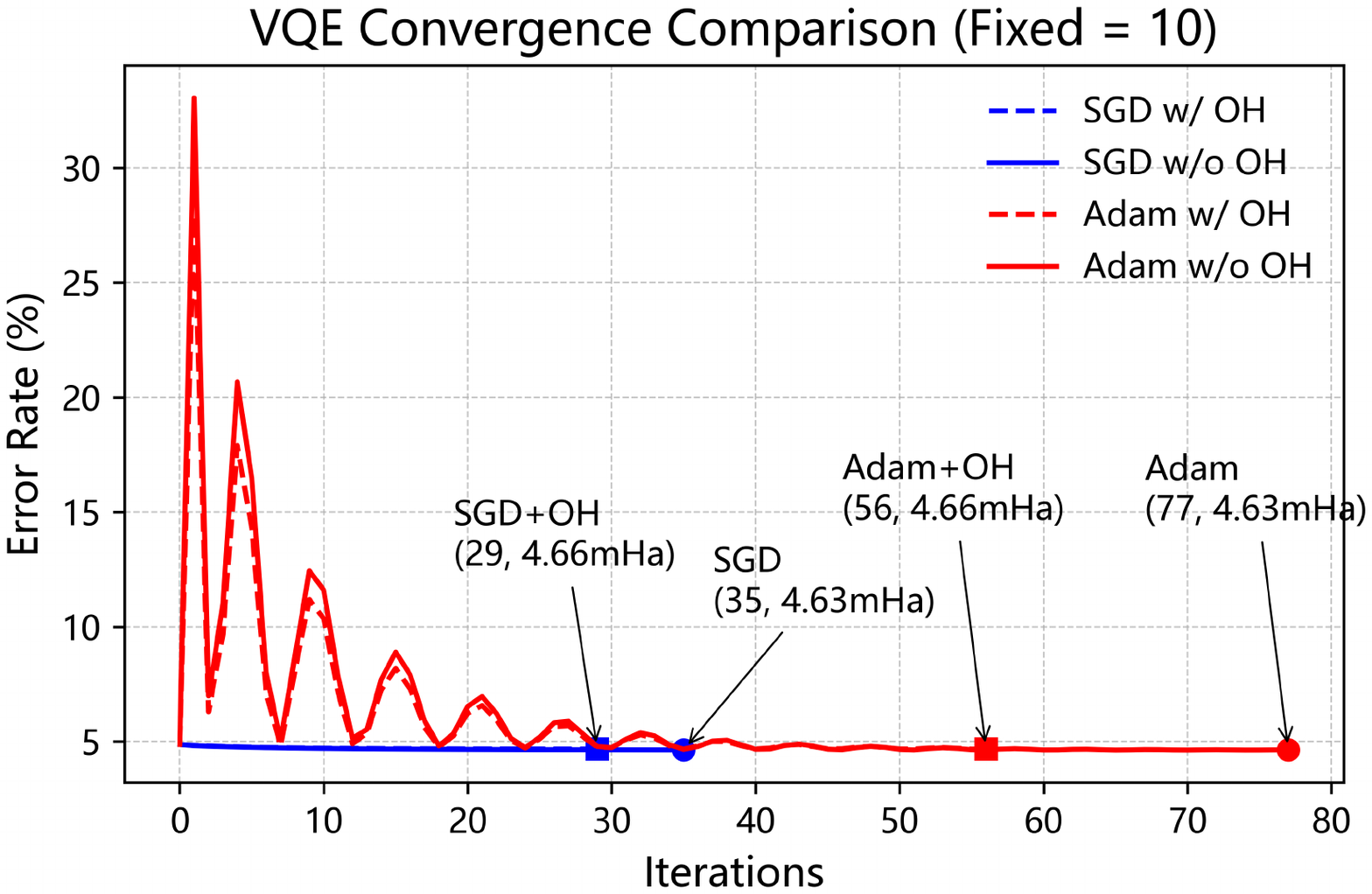} \caption{ Convergence of VQE error rates for different optimizers and training datasets on H$_2$O (fixed parameter size = 10). The error rate is plotted against the number of optimization iterations. Results are shown for both SGD and Adam optimizers, each trained with and without the inclusion of OH$^-$ data. The plot demonstrates that both the choice of optimizer and the presence of OH$^-$ in the training set significantly influence convergence speed and final error rate. Notably, the Adam optimizer with OH$^-$ data reaches the target error threshold in fewer iterations than other configurations. } \label{fig:H2O} \end{figure}

In Figure~\ref{fig:H2O}, we evaluate the accuracy of the predicted ground-state energy of the water molecule. The reference energy used in our study is the Full Configuration Interaction (FCI) result, given by $E_{\mathrm{FCI}} = -75.0116 \ \mathrm{Ha}$. We compare the performance of the LSTM-VQE-Dense model when trained with and without the inclusion of OH\textsuperscript{--} in the training set.

The results demonstrate that incorporating OH\textsuperscript{--} data during training leads to a notable reduction in the number of optimization iterations required to reach convergence. This indicates that the LSTM model has successfully learned optimization patterns from the VQE trajectory of the OH\textsuperscript{--} molecule. Furthermore, the final predicted energy achieves an error within 50 mHa of the FCI reference value.

To further understand how model capacity influences performance, we examine the effect of LSTM parameter size across different molecules.

Table~\ref{tab:h2o_h4_param_compare} summarizes the performance of the LSTM-VQE model with varying parameter sizes for two molecules: H\textsubscript{2}O and H\textsubscript{4}. The table reports the number of iterations required for convergence and the final error rates for models trained using both SGD and Adam optimizers.

For H\textsubscript{2}O, the best performance is observed when the parameter size is 20, achieving error rates of 41.98 mHa with Adam and 44.93 mHa with SGD. However, increasing the parameter size to 40 leads to a substantial degradation in performance with SGD, as the error rate escalates to 676.46 mHa, suggesting potential overfitting or instability. In contrast, Adam maintains robustness, achieving a relatively low error of 38.75 mHa even at the larger parameter size.

For H\textsubscript{4}, performance improves consistently with increasing model capacity. The error decreases from 21.73 mHa at parameter size 10 to within chemical accuracy (0.05 mHa, where chemical accuracy is defined as 1.6 mHa) at size 40 when using the Adam optimizer. This trend indicates that higher-capacity models are more expressive and particularly effective for simpler systems like H\textsubscript{4}.

Notably, the superior accuracy observed on H\textsubscript{4} may also stem from the fact that VQE itself performs well on this molecule. For more complex systems such as H\textsubscript{2}O, the standard VQE typically converges to an energy of approximately $-74.97$ Ha, as reported in \cite{10947806}. Their approach—using a hardware-efficient ansatz (HEA), the UCCSD ansatz, and the COBYLA optimizer—required over 2500 iterations to reach this energy level. In contrast, our fully connected LSTM-VQE model achieves a comparable VQE-quality baseline in just 78 iterations, demonstrating significantly improved computational efficiency. These results indicate that while our model dramatically accelerates convergence, achieving further improvements in absolute accuracy for complex molecules may require enhancements to the underlying VQE framework itself.

\begin{table}[H]
\centering
\caption{Comparison of LSTM-VQE error rates (mHa) and convergence iterations on H\textsubscript{2}O and H\textsubscript{4} under different parameter sizes. For H\textsubscript{2}O, models are trained with OH$^{-}$ included in the training set.}
\label{tab:h2o_h4_param_compare}
\resizebox{\columnwidth}{!}{%
\begin{tabular}{c|c|cc|cc}
\toprule
Molecule & Param Num & \multicolumn{2}{c|}{SGD} & \multicolumn{2}{c}{Adam} \\
         &           & Iterations & Error (mHa) & Iterations & Error (mHa) \\
\midrule
\multirow{4}{*}{H\textsubscript{2}O} 
         & 10 & 29  & 46.63 & 56  & 46.60 \\
         & 15 & 31  & 46.15 & 72  & 44.14 \\
         & 20 & 65  & 44.93 & 70  & 41.98 \\
         & 40 & 120 & 676.46 & 77  & 37.75 \\
\midrule
\multirow{7}{*}{H\textsubscript{4}} 
         & 10 & $\geq$ 501 & 21.92 & 63  & 21.73 \\
         & 15 & $\geq$ 501 & 15.33 & 43  & 15.58 \\
         & 20 & 438 & 4.64  & 71  & 4.53 \\
         & 25 & 343 & 7.92  & 64  & 7.90 \\
         & 30 & 282 & 2.03  & 77  & 1.90 \\
         & 35 & $\geq$ 501 & 1.73 & 60  & 1.22 \\
         & 40 & 333 & 0.16  & 86  & 0.05 \\
\bottomrule
\end{tabular}%
} 
\end{table}



\section{Conclusion}

In this work, we investigate the use of LSTM-based machine learning techniques to accelerate the optimization process of variational quantum eigensolver (VQE) algorithms in molecular energy simulations. Building upon the "learn-to-learn" meta-learning paradigm introduced by PennyLane, we adapt this approach to the domain of quantum chemistry. Our analysis reveals that the original LSTM-VQE framework is limited in its general applicability to diverse molecular systems, primarily due to the rigid structure of wavefunction parameters, which are constrained by molecule-specific qubit counts and Ansatz architectures.

To address this limitation, we propose an enhanced model, \textbf{LSTM-FC-VQE}, which integrates a fully connected layer to project variable-length input features into a unified parameter space. This design enables the model to handle molecular systems with heterogeneous parameter dimensions more effectively.

We validate our approach through experiments on both the simple H\textsubscript{4} molecule and the more complex H\textsubscript{2}O system. The results demonstrate that LSTM-FC-VQE, when combined with the Adam optimizer and a carefully chosen learning rate schedule, significantly reduces the number of VQE iterations required for convergence while maintaining high energy accuracy. For H\textsubscript{4}, the model successfully reconstructs the full energy curve in agreement with full configuration interaction (FCI) results. For H\textsubscript{2}O, it achieves convergence to energy levels comparable to those obtained by conventional VQE methods.

Importantly, our method achieves these results with substantially reduced computational cost. The LSTM-based model learns the iterative optimization behavior of VQE, enabling it to predict high-quality initial parameters that lead to faster convergence without compromising accuracy.

We also examine the generalization capacity of our approach and show that it can extrapolate to previously unseen molecular configurations with minimal training data. Notably, increasing the LSTM network’s parameter size enhances performance on simpler systems but introduces a risk of overfitting in more complex molecules. In such cases, the Adam optimizer proves effective in mitigating overfitting and maintaining stable convergence.

In summary, the LSTM-FC-VQE framework offers a scalable, data-efficient, and optimizer-friendly meta-learning strategy for accelerating variational quantum simulations. By leveraging prior molecular optimization trajectories, the model generalizes well across different systems, delivering high-quality parameter initializations. Future directions include incorporating more expressive sequence models such as attention-based architectures and exploring cross-molecular transfer learning to further enhance generalization in larger and more diverse quantum systems.


\section{Future Work}

Looking ahead, there are several promising directions to further enhance the capabilities and impact of the LSTM-FC-VQE framework in quantum computational chemistry. While our approach demonstrates strong performance on small- to medium-sized molecular systems, extending its applicability to larger and more strongly correlated molecules remains a significant challenge. Future research could explore the integration of graph neural networks (GNNs) \cite{gilmer2017neural} or transformer-based architectures \cite{schwaller2019molecular} to encode richer structural information and capture long-range quantum correlations more effectively.

Incorporating domain knowledge—such as molecular symmetries, physical constraints, or chemically informed feature engineering—into the neural architecture may further improve both generalization and interpretability. A particularly promising direction involves the development of adaptive, data-driven Ansatz selection strategies. Such mechanisms would allow the meta-learning model to recommend not only parameter initializations but also suitable circuit architectures tailored to specific molecular tasks \cite{zhang2023meta}.

From an algorithmic standpoint, combining the LSTM-FC-VQE approach with active learning \cite{smith2018less} or reinforcement learning \cite{zhou2017optimizing} frameworks could enable autonomous exploration of chemical space and optimization landscapes, thereby reducing the dependence on large, labeled datasets. Additionally, techniques such as cross-molecule transfer learning \cite{ramsundar2015massively} and continual learning could be employed to promote knowledge transfer and improve robustness when encountering novel molecular species or chemical reactions.

\section*{Data Availability}
All code and data used in this study are available at the following GitHub repository: \\
\url{https://github.com/leo07010/LSTM-FC-VQE/tree/main}

The repository includes two implementations of LSTM-enhanced Variational Quantum Eigensolvers (VQE): (1) LSTM-FC-VQE, a novel method that integrates Long Short-Term Memory (LSTM) networks with Fully Connected layers to optimize UCCSD ansatz parameters, and (2) Reference-LSTM-VQE, a baseline method for comparison. The code, dependencies, and usage instructions are provided to enable full reproducibility of the results.

\section*{Acknowledgment}
EJK thanks the National Center for Theoretical Sciences and National Yang Ming Chiao Tung University in Taiwan for their support.

RYC thanks Yang Ming Chun for his assistance during the research process and for providing physical knowledge and verification.

\bibliographystyle{IEEEtran}
\bibliography{references}
\appendix

\section{Model Training Details}

In this appendix, we provide details of the model architecture and training procedures, including the choice of hyperparameters.

We trained two versions of the LSTM-VQE framework to evaluate different model design choices: a baseline LSTM-VQE (vanilla) model and an enhanced LSTM-FC-VQE model that incorporates a fully connected layer. The shared and distinct training configurations for these models are summarized below. All LSTM models were trained using either an NVIDIA RTX 4060 GPU or Google Colab's T4 GPU. The VQE evaluations were performed exclusively on CPU, reflecting a separation between model training and quantum simulation execution.

\subsection{Common Settings}

\begin{itemize}
\item \textbf{Framework:} TensorFlow and PennyLane were used, with circuits executed on the Lightning-QML backend.
\item \textbf{Dataset:} Molecules included H\textsubscript{2}, H\textsubscript{3}$^+$, H\textsubscript{4}, and OH$^-$. Molecular geometries and Hamiltonians were accessed directly from the PennyLane Molecules Dataset using the STO-3G basis\cite{Utkarsh2023Chemistry}.
\item \textbf{Loss function:} Weighted sum of energy over time steps:
\begin{equation}
L = \frac{1}{T}\sum_{t=1}^{T} 0.1 \times t \cdot E_t(\theta)
\end{equation}
\item \textbf{Early stopping:} Training was terminated if relative change in loss fell below $10^{-4}$.
\item \textbf{Training schedule:} The LSTM model was trained using the Adam optimizer. The predicted parameters were then passed to a VQE routine, which was evaluated using both Adam and SGD optimizers, under constant and exponential decay learning rate schedules.
\end{itemize}

\subsection{Vanilla LSTM-VQE}

\begin{itemize}
\item \textbf{Ansatz types:} UCCSD, HEA, and StronglyEntangle were trained separately.
\item \textbf{Model:} A standard LSTMCell with hidden dimension equal to the largest ansatz parameter count. Output is directly interpreted as variational parameters.
\item \textbf{Input:} Each time step received zero-initialized parameters, hidden states, and energy cost. All molecule data, including Hamiltonians, were loaded from PennyLane.

\end{itemize}

\subsection{LSTM-FC-VQE}

\begin{itemize}
\item \textbf{Ansatz:} UCCSD only.
\item \textbf{Model:} One shared LSTM layer (latent dimension 40) followed by two separate fully connected layers for single and double excitation parameters.
\item \textbf{Input:} Each molecule's excitation parameters were flattened into a fixed-length vector. Only the first $M$ components—matching the LSTM input size—were retained as training input.

\end{itemize}

This structure ensures comparability while highlighting the improvements introduced by the FC-layer-enhanced model.

\end{document}